\documentclass[12pt]{article}
\usepackage{graphicx}  
\begin{document}
\baselineskip=24pt
\title{Composite Grains: Effects of Porosity and Inclusions on 
the 10$\mu$m Silicate Feature} 
\author{Deepak. B. Vaidya$^{1}$
and Ranjan Gupta$^2$\thanks{E-mail:rag@iucaa.ernet.in}\\
$^{1}$Gujarat Arts \& Science College, Ahmedabad-380006, India\\
$^{2}$IUCAA, Post Bag 4, Ganeshkhind, Pune-411007, India\\
}

\maketitle
\clearpage
\newpage
\noindent
\newpage
\label{firstpage}

\begin{abstract}
We calculate the absorption efficiency of the composite grains, made up
of host silicate spheroids and inclusions of ices/graphites/or voids, 
in the spectral region  $7.0-14.0\mu$m  The absorption
efficiencies of the composite spheroidal grains for three axial
ratios are computed using the discrete dipole approximation (DDA)
as well as using the effective medium approximation \& T-Matrix (EMT-Tmatrix) approach.
We study the absorption as a function of the volume fraction of the 
inclusions and porosity. In particular, we study the variation
in the $10.0\mu$m feature with the volume fraction of the 
inclusions and porosity. We then calculate the infrared fluxes
for these composite grains and compare the model curves with the 
average observed IRAS-LRS curve, obtained for several circumstellar 
dust shells around stars.
These results on the composite grains show
that the wavelength of the peak absorption
shifts and the width of the $10.0\mu$m feature
varies with the variation in the volume fraction
of the inclusions. The model curves for composite
grains with axial ratios not very large (AR$\sim$1.3)
and volume fractions of inclusions with f=0.20,
and dust temperature of about 250-300$^{\circ}$K ,
fit the observed emission curves reasonably well.
\end{abstract}

%

\newpage
\noindent

\section{Introduction}
Although, the cosmic dust grains are in general not homogeneous spheres,
usually the observed absorption, scattering and extinction data are
interpreted using calculations based on Mie theory, which is strictly
valid for homogeneous spherical particles.
Dust grains ejected from stars are more likely to be non-spherical and
inhomogeneous, viz. porous, fluffy and composites of many small grains glued
together, due to grain-grain collisions, dust-gas interactions and various
other processes. Since there is no exact theory to study the scattering properties
of these inhomogeneous grains, there is a need for
 formulating models of electromagnetic scattering by these grains. 
Mathis \& Whiffen (1989), Mathis (1996), Dwek (1997) and Li \& Greenberg (1998)
have proposed composite
grain models consisting of silicate and amorphous carbon as
constituent materials. They have used effective medium
approximation (EMT) to calculate the optical constants for composite
grains and then used the Mie theory to calculate extinction cross sections
for spheres. In EMT the inhomogeneous particle is replaced by a
homogeneous one with some 'average effective dielectric function'.
The effects related to the fluctuations of the dielectric function within
the inhomogeneous structures cannot be treated by this approach
of the EMT.
Iati etal (2004) have studied optical properties of composite grains
as grain aggregates of amorphous carbon and astronomical silicates,
using the transition matrix approach. 
Recently Voshchinnikov etal (2006) and Voshchinnikov \& Henning (2008)
have studied the effect
of grain porosity on interstellar extinction, dust temperature,
infrared bands and millimeter opacity. They have used both, the
EMT-Mie based calculations and layered sphere model.

Earlier, we have used Discrete Dipole Approximation (DDA) to study
the extinction properties of the composite grains (Vaidya etal 2007).
For the description on the DDA see Draine (1988).
The DDA allows the consideration of irregular shape effects,
surface roughness and internal structure within the grain
(Wolff et al. 1994, 1998 and Voshchinnikov et al. 2005).
For discussion and comparison of DDA and EMT methods, including
the limits of the effective medium theory, see
Bazell and Dwek (1990), Perrin and Lamy (1990), Perrin and Sivan
(1990), Ossenkopf (1991) and Wolff et al (1994).

The paper has the following sections:
In section 2 we give the
validity criteria for the DDA and the composite grain models.
In section 3 we present and discuss the results of our computations and compare
the model curves with the observed IR fluxes obtained by IRAS satellite.
The main conclusions of our study are given in section 4.

\section{Composite grains and DDA} 

We use the computer code developed by Dobbie (1999) to generate the
composite grain models used in the present study.
We have studied composite grain models with a host silicate
spheroid containing N= 9640, 25896 and 14440 dipoles, each carved out from
$32 \times 24 \times 24$, $48 \times 32 \times 32$
and  $48 \times 24 \times 24$ dipole sites, respectively;
sites outside the spheroid are set to be vacuum (void) and sites inside are
assigned to be the host material.
It is to be noted that the composite spheroidal grain with
N=9640 has an axial ratio of 1.33, whereas N=25896 has the axial
ratio 1.5 and N=14440 has the axial ratio 2.0.
The volume fractions of the graphite inclusions used are
10\%, 20\% and 30\% (denoted as f=0.1, 0.2 and 0.3)
Details on the computer code and the corresponding modification
to the DDSCAT code (Draine \& Flatau 2003) are given in Dobbie
(1999), Vaidya et al. (2001, 2007) and Gupta et al. (2006).
The modified code outputs a three-dimensional matrix specifying
the material type at each dipole site; the sites are either silicate,
graphite or vacuum (void).
For an illustrative example of a composite spheroidal grain
with N=9640 dipoles, please see Figure 1 (also Vaidya et al, 2007).
There are two validity criteria for DDA (see e.g. Wolff et al. 1994);
viz. (i) $\rm |m|kd \leq 1$, where m is the complex refractive index
of the material, k=$\rm 2\pi/\lambda$ is the wavenumber and
d is the lattice dispersion spacing and
(ii) d should be small enough (N should be sufficiently large) to
 describe the shape of the particle satisfactorily.
The complex refractive indices
for silicates and graphite are obtained from Draine (1985, 1987) and
that for ice is from (Irvine \& Pollack 1968) .
For any grain model, the number of dipoles required to
obtain a reliable computational result can be estimated using the
DDSCAT code (see Vaidya and Gupta 1997 and 1999, Vaidya etal 2001).
For the composite grain model, if the host grain has N dipoles,
its volume is N(d)$^3$ and if 'a' is the
radius of the host grain , N(d)$^3$ =4/3$\rm \pi(a)^3$,
hence, N = 4$\rm \pi/3(a/d)^3$, and if $\rm |m|kd$=1 and
k=2$\rm \pi/\lambda$
the number of dipoles N can be estimated at a given wavelength
and the radius of the host grain.

It must be noted here that the composite spheroidal grain
models with N=9640, 25896 and 14440 have the axial ratio 1.33,
1.5 and 2.0 respectively and if the semi-major axis and semi-minor 
axis are denoted by x/2 and y/2
respectively, then
$a^3=(x/2)*(y/2)^2$, where 'a' is the
radius of the sphere whose volume is the same as that of
a spheroid.
In order to study randomly oriented spheroidal grains,it is
necessary to get the scattering properties of the composite
grains averaged over all of the possible orientations; in the
present study we use three values for each of the orientation
parameters ($\beta, \theta \& \phi$),i.e. averaging over 27 orientations,
which we find quite adequate (see e.g. Wolff etal 1998).  

\section{Results and Discussion}

\subsection{Absorption Efficiency of Composite Spheroidal Grains}

Earlier, we had studied the extinction properties of composite
spheroidal grains in the spectral region 3.4-0.10$\mu$m
(Vaidya etal 2007).
In the present paper, we study the absorption properties of the
composite spheroidal grains
with three axial ratios, viz. 1.33, 1.5 and 2.0, corresponding to
the grain models with N=9640, 25896 and 14440 respectively,
in the wavelength region 7.0-14.0$\mu$m. The inclusions selected
are graphites/ices/or voids. 

Figure 2 (a-c) shows the Absorption efficiencies
($\rm Q_{abs}$) for the
composite grains with the host silicate spheroids containing 9640,
25896 and 14440 dipoles, corresponding to axial ratio 1.33, 1.5
and 2.0 respectively with a power law size distribution of a range of
grain sizes from 0.005-0250$\mu$m in steps of 0.005$\mu$m.
The three volume fractions, viz. 10\%, 20\% and 30\%, of ice
inclusions are also listed in the top (a) panel.

The effect of the variation of volume fraction of inclusions is
clearly seen for all the models.
It is to be noted that the wavelength of the peak absorption
shifts with the variation in the volume fraction of inclusions.
These absorption curves also show the variation in the width
of the absorption feature with the volume fraction of inclusions.
All these results indicate that the inhomogeneities within the
grains play an important role in modifying the 10$\mu$m feature.
It is also seen in Fig. 2 that the shape of the
Q$_{abs}$ curve also varies with the axial ratio,
AR of the composite grain, and the absorption  
peak shifts towards shorter wavelength as the
AR incereases.

O'Donnell (1994) has investigated the variation of the 10$\mu$m
feature with the grain composition using spheroidal composite
grains containing inclusions of either amorphous carbon, tholins
or vacuum. He found that the inclusion of tholins or carbon 
increases the absorption on the short wavelength side of the
10$\mu$m feature. Further, the inclusion of tholin also 
broadens the 10$\mu$m feature.

Figure 3 shows the absorption efficiency for the composite grains
(N=9640, AR=1.33, a=0.10$\mu$m) with host silicate and
graphite inclusions (thick lines). It is seen that the Q$_{abs}$ decreases as the
the volume fraction of inclusions ' f ' increases and the
absorption peak shifts towards shorter wavelength and
broadens with the volume fraction of inclusions. 
We have compared our results on the absorption efficiencies
of the composite grains obtained using the DDA with the results obtained
using the EMT-T matrix based calculations. These results are
also displayed in Figure 3 (in thin lines). For these calculations, the optical constants
were obtained using the Maxwell-Garnet mixing rule 
(Bohren and Huffman 1983). Description of the T-matrix method/and code is
given by Mishchenko (2002).
 
Since Maxwell-Garnett rule provides the extreme cases of the
possible values of the effective dielectric constants of the 
two-component mixture (see Chylek et al, 2000), we have used this 
rule. On the other hand, Bruggeman mixing rule
applies to a two-component mixture with no distinguishable
inclusions embedded in a definite matrix; i.e. it
applies to a completely randomly inhomogeneous medium
(Bohren \& Hoffman, 1983).
In the figure 3 we have also shown the absorption efficiency 
for the composite grain with f=0.2 using Bruggeman rule. 

It is seen that the absorption curves obtained using
EMTs (Maxwell-Garnett and Bruggeman mixing
rules) deviate considerably from the curves obtained
using DDA. The DDA allows consideration of irregular
shape effects, surface roughness and internal structure
within the grain (Wolff et al 1994, 1998). In EMT the 
inhomogeneous grain is replaced by a homogeneous
one with some 'average effective dielectric function',
the effects related to the fluctuations of the dielectric
function within the inhomogeneous structures can not
be treated by the EMTs and material interfaces and
shapes are smeared out into a homogeneous
'average mixture' Saija etal 2001). Perrin and Lamy (1990)
have calculated extinction cross sections for the composite
grains, using both the EMTs, as well as the DDA.
They found that for small porous grains the results given by
Maxwell-Garnett are much better than the results given by
the Bruggeman theory. The criteria of validity
of the respective effective theories are not clear (Perrin and
Lamy, 1990). We have used more accurate DDA method to
calculate the absorption cross sections for the composite grains. 
However, it would still be very useful and desirable
to compare the DDA results for the composite grains with those computed by other
EMT/Mie type/T matrix techniques in order to examine the applicability
of several mixing rules. (see Wolff etal 1998, Voshchinnikov and Mathis
1999, Chylek et al. 2000, Voshchinnikov etal 2005, 2006). 
The application of DDA, poses a computational
challenge, particularly for the large values of
the size parameter $X (=2\pi a/\lambda > 20 )$ and the complex refractive index m of
of the grain material would require large number of dipoles and that in
turn would require considerable computer memory and cpu time (see e.g. Saija etal
2001, Voshchinnikov etal 2006). 

We have also calculated the absorption efficiencies of the porous silicate grains.
Figure 4 shows the absorption efficiencies of the composite grains (N=9640) with
the host silicate spheroids and voids as inclusions in three
volume fractions viz. 0.05, 0.10 and 0.20 for the 
three grain sizes viz. a=0.1, 0,2 and 0.5$\mu$m. 
These results show that the peak strength decreases with the 
porosity. It is also seen that the peak of the 10$\mu$m shifts
and broadens as the porosity increases.
Voshchinnikov et al (2006) and Voshchinnikov \& Henning (2008)
have used layered spheres model combining all components
including vacuum, to study the effect of porosity on the
10$\mu$m feature. Voshchinnikov \& Henning (2008) found 
the peak strength decreases and the feature also broadens
with increasing porosity. These results are consistent
with the results we have obtained with our composite grain
model.
Recently, Li et al (2008) have also used the porous grains 
to model the 9.7$\mu$m feature in the AGN.
They have used multilayered sphere model of Voshchinnikov \&
Mathis (1999) and calculated the absorption efficiency
of porous composite dust consisting amorphous silicate,
amorphous carbon and vacuum. Li et al (2008) have found
the progressive broadening of the 10$\mu$m feature and shift
to longer wavelengths of the peak position of the feature
as the porosity increases. 

\subsection{Infrared Emission from Circumstellar Dust:
    Silicate Feature at 9.7$\mu$m}

In general, stars which have evolved off the main sequence and
which have entered the giant phase of their evolution are a major
source of dust grains in the galactic interstellar medium. Such
stars have oxygen overabundant relative to carbon and therefore
produce silicate dust and show the strong feature
at 9.7$\mu$m.  This is ascribed to the Si-O stretching mode in
some form of silicate material, e.g. olivine. These materials
also have a feature at 18$\mu$m, resulting from the 
O-Si-O bending mode (Evans, 1993).

Using the absorption efficiencies of the composite grains, we calculate
the infrared flux, F$_{\lambda}$ at various temperatures of the dust,
and for a power law MRN dust grain size distribution (Mathis etal 1977). 
We compare the model curves with the average observed LRS-IRAS curve,
obtained for several circumstellar dust shells around stars.
Figure 5(a) shows the infrared flux at the dust temperature T=250$^{\circ}$K
for the composite grains containing number of dipoles N=9640,
for three volume fractions of void inclusions (porous).
Figure 5(b) shows infrared flux at T=450$^{\circ}$K for the 
composite grains with voids as inclusions.

Figure 6 shows the variation of infrared flux between T=200 and 500$^{\circ}$K, for the
porous grains. In this figure we have also shown the average observed IRAS-LRS
curve (Whittet 2003).

These results show that the composite grains with number of
dipoles, N=9640 and porosity f=0.2,
and dust temperature around 300$^{\circ}$K fit the
observed infrared flux (Whittet, 2003) reasonably well. It is to be noted here that
Voshchinnikov \& Henning (2008) have compared the results of the porous
composite grains with the spectra of T Tauri and Herbig Ae/Be stars,
Li et al (2008) have compared the porous grain models with the AGNs, whereas
in this paper we have compared the composite grain models with the average
observed IRAS flux (Whittet, 2003).

\section{Summary and Conclusions}

Using the discrete dipole approximation (DDA)
we have studied the variation of the absorption efficiency for the composite
spheroidal grains, with the volume fractions of the inclusions
in the wavelength region of 7.0-14.0$\mu$m.
These results clearly show the variation in the absorption efficiency for the composite
grains with the volume fractions of the inclusions. The results on the composite
grains also show the shift in the peak absorption wavelength 9.7$\mu$m
with the volume fraction of the inclusions and porosity.
The results obtained using DDA based calculations deviate considerably from
the results obtained using EMT-TMatrix calculations.

The results on the composite grains clearly show that the
inhomogeneity in the grains modifies the absorption/emission
properties of the grains. Further study on the emission
properties
of composite grains at longer wavelengths in the IR is in
progress (Vaidya etal 2008).

\section{Acknowledgments}

DBV and RG thank ISRO-RESPOND for the grant (No. ISRO/RES/2/345/
2007-08) , under which this study has been carried out.
The authors also thank the organizers of ELS-XI conference for their
generous support towards attending the conference.

\clearpage
\noindent
{\Large\bf References}\\

Bazell D. and Dwek,1990,ApJ,360,342

Bohren C. F. and Huffman D. R., 1983, in Absorption and
Scattering of light , Wiley, N.Y, p. 217

Chylek, P., Videen, G., Geldart, D.J.W., Dobbie, J., William, H.C.,
2000: in Light Scattering by Non-spherical Particles, Mishchenko,
M., Hovenier, J.W.  and Travis, L.D. (eds), Academic Press, New York, p. 274

 Dobbie J., 1999, PhD Thesis, Dalhousie University,

 Draine B.T., 1985, ApJS, 57, 587

 Draine B.T., 1987, Preprint Princeton Observatory, No. 213

 Draine B.T., 1988, ApJ, 333, 848

 Draine B.T. and Flatau P.J., 2003, DDA code version 'ddscat6'

 Dwek E., 1997, ApJ, 484, 779

Evans A., 1993, in The Dusty Universe, Ellis Horwood Ltd,
England 

Gupta R., Vaidya D.B., Dobbie J.S.,Chylek P., 2006, As.
Sp.Sc.,391,21

Iati, M.A., Giusto, A., Saija, R., Borghese, F.,
Denti, P., Cecchi-Pestellini, C. and Aielo, S., (2004), ApJ, 615, 286

Irvine W. M. and Pollack J. B., 1968, ICARUS, 8,324

Li A., and Greenberg, J.M., 1998, A\&A, 331, 291

 Li M.P., Shi Q.J.,and Li, A.., 2008, MNRAS, 390, 778

 Mathis J.S., 1996, Ap.J., 472, 643

 Mathis J.S. and Whiffen G., 1989, Ap J., 341, 808

 Mathis J.S., Rumpl W., Nordsieck K.H., 1977, Ap.J., 217, 425

Mishchenko, M.L., Travis, L.D. and Lacis, A.A., 2002, in
Scattering, Absorption and Emission of Light by Small Particles, CUP,
Cambridge, UK, p.184

O'Donnell, J.E., 1994, ApJ, 437, 262

 Ossenkopf V., 1991, A\&A, 251, 210

 Perrin J.M. and Lamy P.L., 1990, ApJ, 364, 146

 Perrin J.M. and Sivan J.P. 1990, A \& A, 228, 238

 Saija R.,Iati M.,Borghese F.,Denti P.,Aiello S.,Cecchi-Pestellini C.  ApJ.,(2001), 539,993

 Vaidya D.B. and Gupta R., 1997, A \&A, 328, 634

 Vaidya D.B. and Gupta Ranjan, 1999, A \& A, 348, 594

 Vaidya D.B., Gupta R., Dobbie J.S and Chylek P., 2001, A \& A, 375, 584

 Vaidya D.B., Ranjan Gupta and Snow T.P., 2007, MNRAS 379, 791

 Vaidya D.B. and Gupta, R, 2008, In preparation.

Voshchinnikov N.V. and Mathis, J.S., 1999, ApJ, 526, 257

Voshchinnikov N.V. and Henning Th., 2008, A\& A, L9, 483

 Voshchinnikov N. V., 2002, in Optics of Cosmic Dust,
 eds. Videen G. and Kocifaj M., Kluwer, 3\\

 Voshchinnikov, N.V., Il'in, V.B. and Th. Henning, 2005,
 A\&A, 429, 371

 Voshchinnikov N.V.,Il'in V.B.,Henning Th.,Dobkova D.N.. 2006,
A\&A,445,993

 Whittet D.C.B., 2003, in Dust in the Galactic Environments,
pp 125, 2nd Edn. (IOC Publishing Ltd. UK)\\

 Wolff M.J., Clayton G.C., Martin P.G. and Sculte-Ladlback
 R.E., 1994, ApJ, 423, 412

 Wolff M.J., Clayton G.C. and Gibson S.J., 1998, ApJ,
503, 815

\clearpage
\newpage
\noindent
\begin{figure}
\includegraphics[width=140mm]{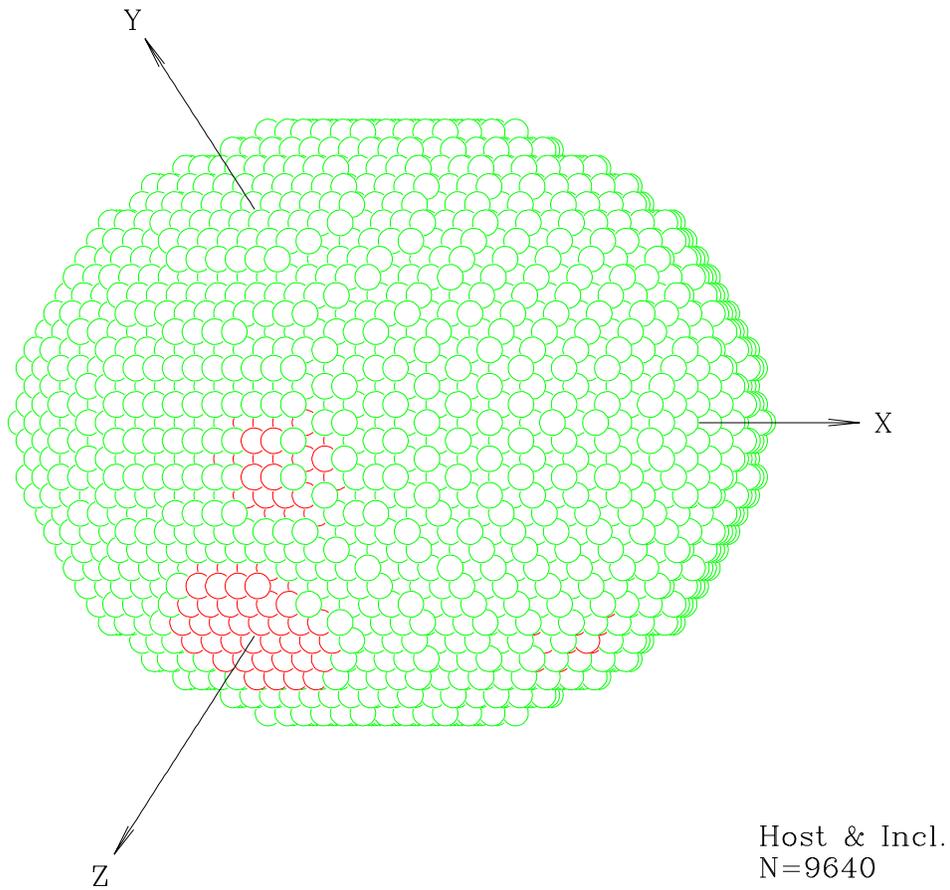}
\caption{A typical non-spherical composite grain with a
total of N=9640 dipoles where the inclusions
embedded in the host spheroid are shown such that only the ones placed
at the outer periphery are seen (Reproduced from Vaidya et al, 2007).}
\end{figure}

\begin{figure}
\includegraphics[width=140mm]{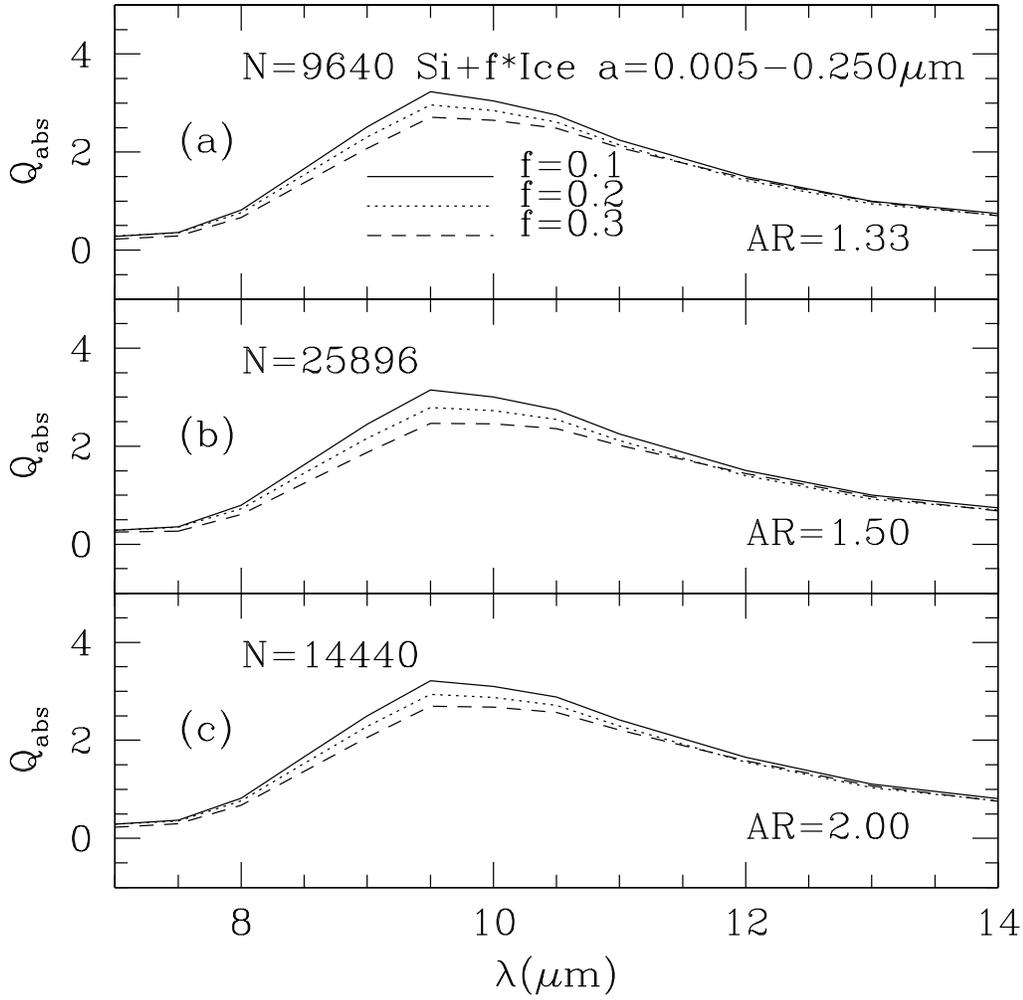}
\caption{Absorption Efficiencies for the composite grains
 with host silicate spheroids containing N=9640; 25896 \& 14440 dipoles and
ices as inclusions for three volume fractions.}
\end{figure}

\begin{figure}
\includegraphics[width=140mm]{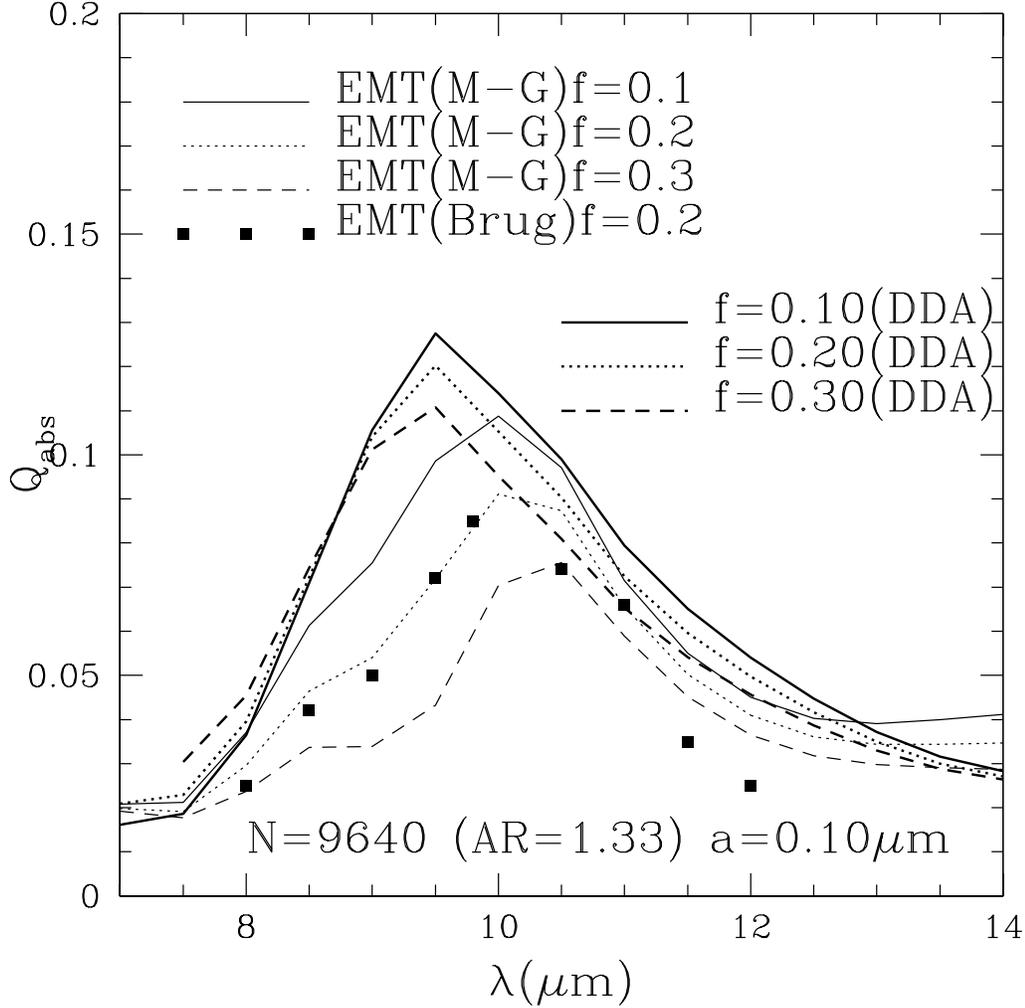}
\caption{Absorption efficiencies for the composite grains, with
host silicate spheroids containing N=9640 dipoles and graphite as inclusions for three
volume fractions for a grain size of a=0.10$\mu$m (thick lines). 
Also shown are the EMT-T matrix calculations 
with the same axial ration AR=1.33 and three volume fractions (thin lines).
The dots represent Bruggeman rule calculations for f=0.2.}
\end{figure}

\begin{figure}
\includegraphics[width=140mm]{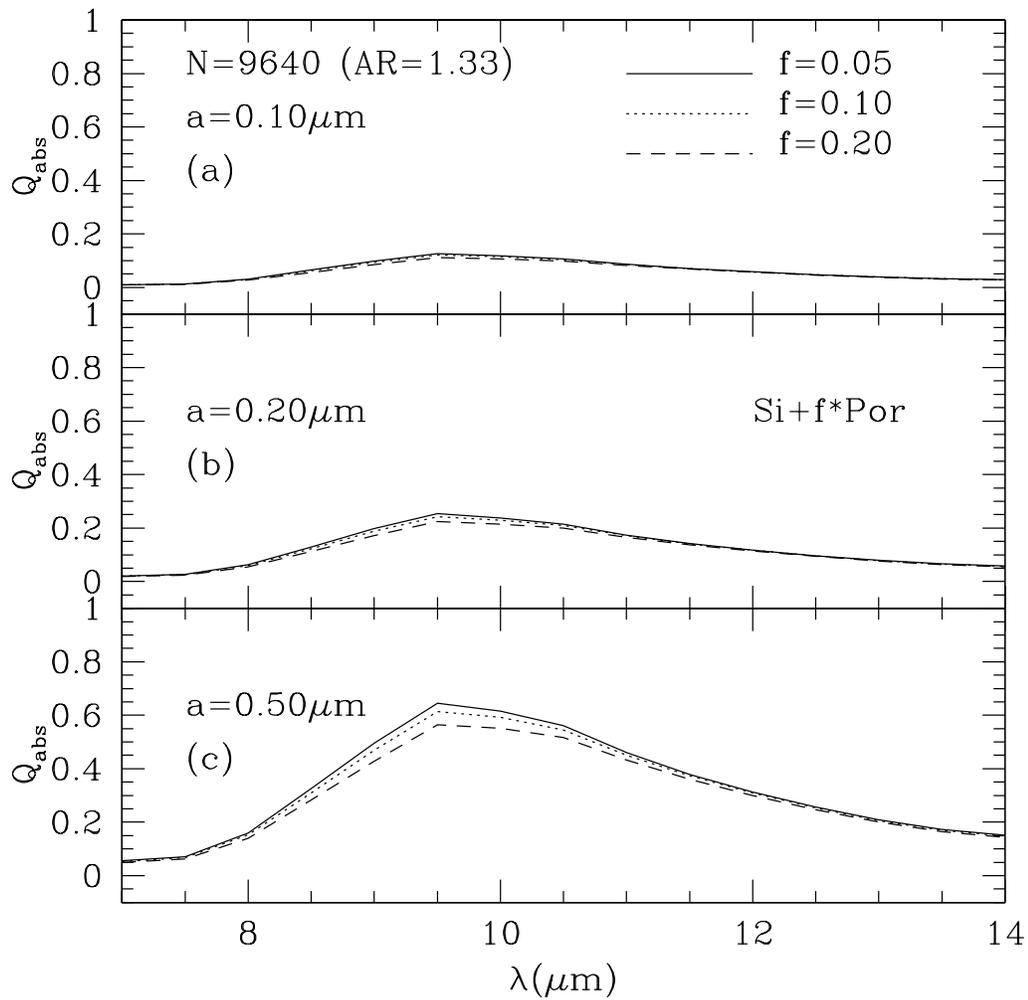}
\caption{Absorption Efficiencies for the porous silicate grains with N=9640, three
volume fractions and three grain sizes.}
\end{figure}

\begin{figure}
\includegraphics[width=140mm]{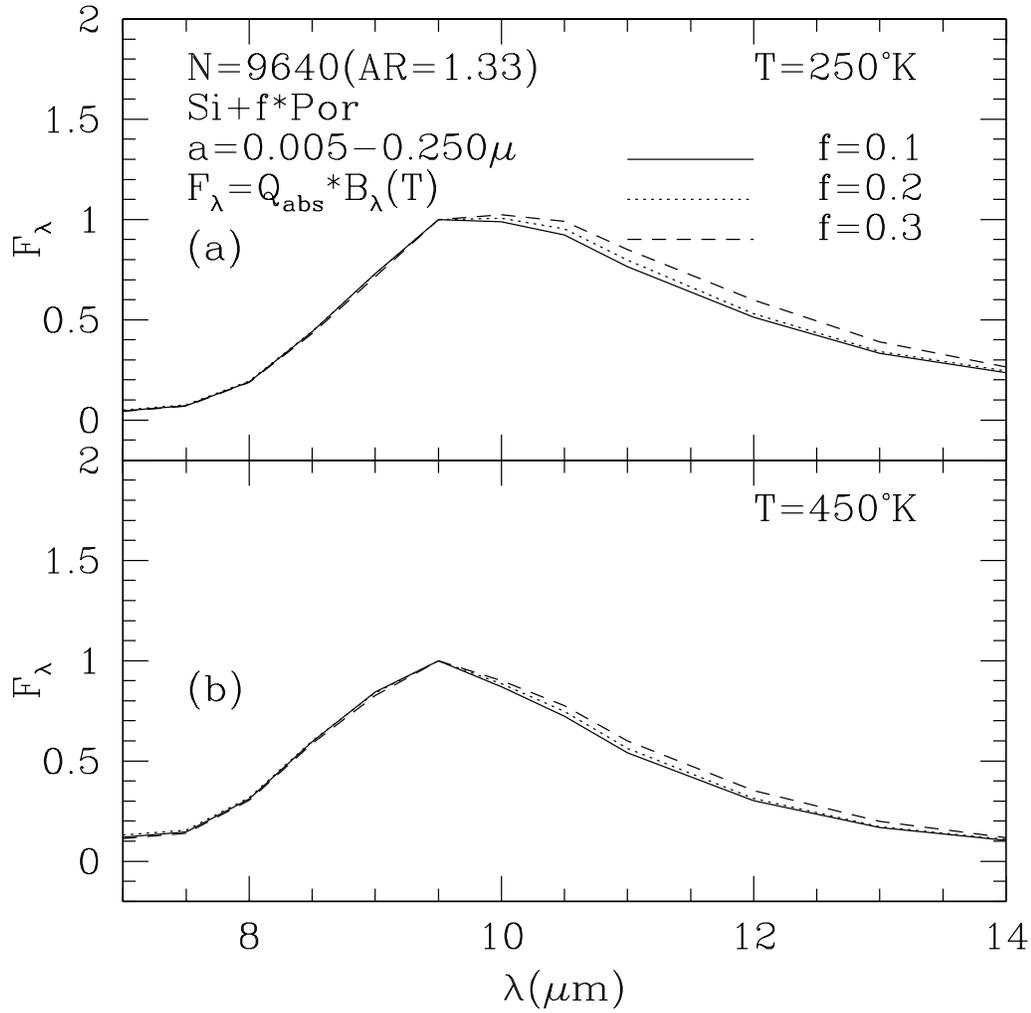}
\caption{Infrared flux for the composite grains with voids as inclusions (porous),
 in the wavelength range of 7-14$\mu$m., at the dust temperatures of 
T=250 \& 450 $^{\circ}$K in panels (a) and (b) respectively.}
\end{figure}

\begin{figure}
\includegraphics[width=140mm]{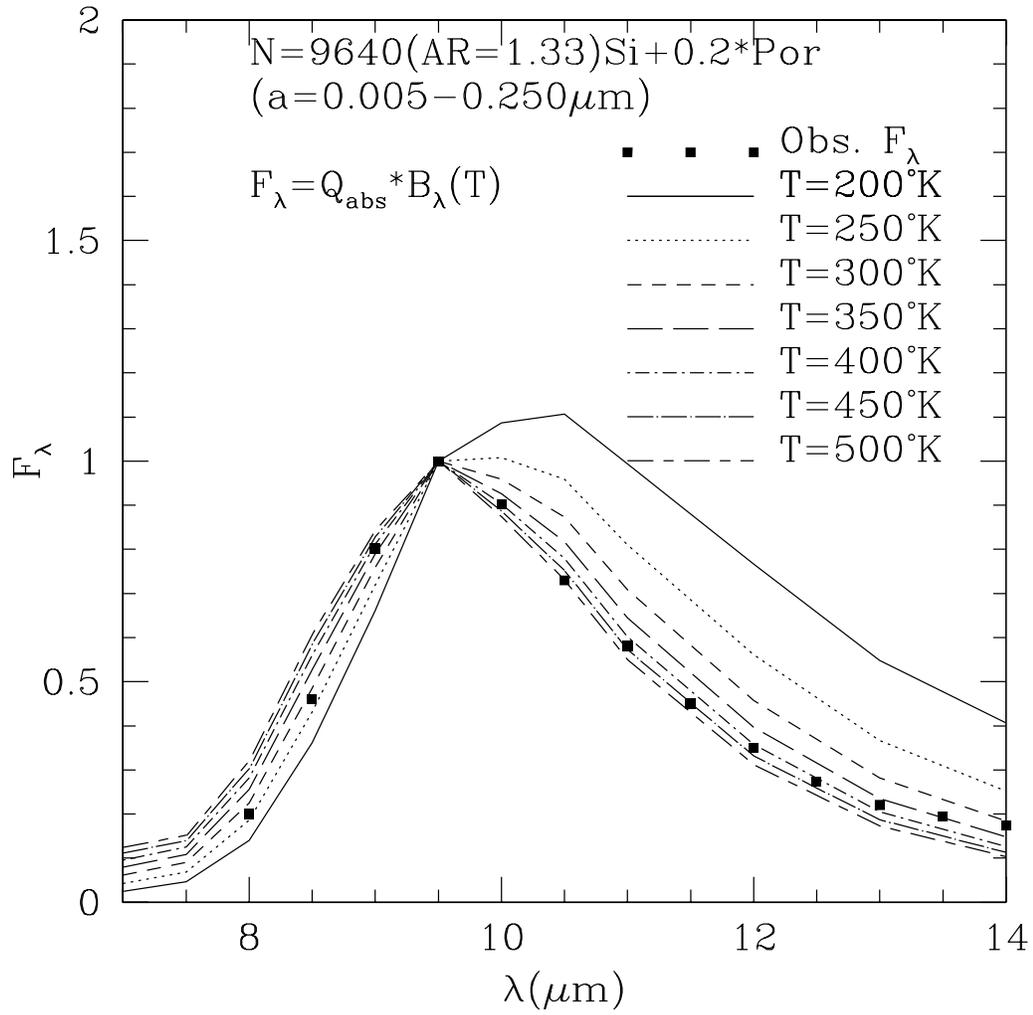}
\caption{Infrared flux for the composite grains with voids as inclusions at various
dust temperatures. The average observed IRAS IR flux values are also shown (in square
dots) for comparison.}
\end{figure}

\end{document}